# Interdisciplinarity and research on local issues: evidence from a developing country

Diego Chavarro[1], Puay Tang[2] and Ismael Rafols[3]

[1] *diego.chavarro@sussex.ac.uk*
SPRU - Science and Technology Policy Research University of Sussex, Brighton (England)

[2] *p.tang@sussex.ac.uk*
SPRU - Science and Technology Policy Research University of Sussex, Brighton (England)

[3] *i.rafols@ingenio.upv.es*
*Ingenio (CSIC-UPV),* Universitat Politècnica de València, València (Spain) & SPRU - Science and Technology Policy Research University of Sussex, Brighton (England)

## Abstract

This paper explores the relationship between interdisciplinarity and research pertaining to local issues. Using Colombian publications from 1991 until 2011 in the Web of Science, we investigate the relationship between the degree of interdisciplinarity and the local orientation of the articles. We find that a higher degree of interdisciplinarity in a publication is associated with a greater emphasis on Colombian issues. In particular, our results suggest that research that combines cognitively disparate disciplines, what we refer to as distal interdisciplinarity, tends to be associated with more local focus of research. We discuss the implications of these results in the context of policies aiming to foster the local socio-economic impact of research in developing countries.

[2]Corresponding author: Puay Tang (p.tang@sussex.ac.uk)

## 1. Introduction

Research in developing countries is often perceived as being overly driven by international agendas and paying insufficient attention to their specific social needs, for example in tropical diseases. (Kreimer, 2007; Thomas, 2010, p. 39). This perception has led to calls for research in developing countries, which now produce a sizeable part of the world's scientific output, to conduct research that addresses their 'mounting societal problems' in areas such as water, food, health, energy and climate change, for example (Macilwain, 2014). Research that in principle matters most for addressing social needs, has long been associated with interdisciplinary research (IDR) (Barry et al., 2008). In this exploratory study, we pose the question whether local-issue research is also most often conducted by interdisciplinary means.

It is widely assumed that research addressing social and economic needs is most often and best conducted through interdisciplinary approaches (Rhoten and Parker, 2006). The perception of the benefits of interdisciplinary research has stimulated a steadily growing interest in developing new knowledge through research that integrates the skills and perspectives of multiple disciplines. The heightened growth of such research

> [may] be in part a parallel of the wider societal interest in holistic perspectives that do not reduce human experience to a single dimension of descriptors, and to awareness that a number of extremely important and productive fields of study are themselves interdisciplinary: biochemistry, biophysics, social psychology, geophysics, informatics… (Aboelela et al. 2007, p. 330).

This article aims to add to the body of literature on the role of IDR to address complex social, cultural, economic and political issues by empirically examining the relationship between IDR and the production of research that addresses local issues. For the purposes of this article, we introduce the term "local issue research" to mean research related to either local, regional or national contexts, conditions or topics, as opposed to research that is universalistic or decontextualised. Following Ordóñez-Matamoros et al. (2010, p. 421), we view local-issue research as research that contributes 'to the local stock of information necessary to increase local understanding and to produce new knowledge valuable to solve local intellectual, technical, or social issues.'

The central hypothesis of this article is that local-issue research tends to be more interdisciplinary than non-local research. We choose Colombia as the exploratory case to test this hypothesis, since it is one of the mid-income countries with a rapidly growing scientific production and which has science policies allegedly supporting IDR as a way to sustain social relevance.

The structure of the paper proceeds as follows. First, we review the arguments provided by the literature suggesting relationships between IDR, social-relevance and local-issue research. Then we present an overview of the policy context for IDR, in particular that of Colombia. Section four explains the operationalization of the key concepts. Section five describes the bibliometric data and methods: the measures of interdisciplinarity and local-issue research, and the logistic regression model. Section six presents the results. Section seven discusses the results and preliminarily explores the policy implications of these findings. We make available to the readers the original data, results of the analysis and computational procedures in three Supplementary Files.[1]

## 2. The relationship between research on local issues and IDR

*Relationship between social relevance and interdisciplinarity*

Scholars have long recognized that IDR is more able to respond to pressing societal questions that pose particular problems (Lowe and Phillipson, 2006; Nightingale and Scott, 2007). For instance, health may not be adequately studied through a single disciplinary framework. Instead, poor health results from a constellation of factors: malnutrition, bad eating habits, genetics, age, poverty, ignorance, pollution, environmental conditions, and peer pressure (for instance, in anorexia). There may be some cases of socially relevant questions that can be answered by monodisciplinary approaches (e.g. in labour economics: is there a relationship between minimum wage and the rate of unemployment?), but these are arguably the exception rather than the rule.

---

[1] Supplementary Files are also available at www.interdisciplinaryscience.net/pub_docs/idr-local-files/

Insights on the relation between IDR and social relevance have been substantiated by recent quantitative studies. There are diverse bodies of literature on social or cognitive diversity in groups or in network relations, which have shown a positive relationship between such diversity and problem-solving and/or creativity outcomes (e.g. Page, 2007, Fleming et al, 2007). D'Este et al. found that researchers with disciplinary diversity are more likely to 'exploit their technology inventions and produce saleable goods and services' (2012, p. 301). In a separate study D'Este et al. (2013) also concluded that cognitive diversity is associated with 'pro-social' research behaviour, that is, attitudes that explicitly take into account the social relevance as an important goal of research. In studies specifically about IDR, Rijnsover and Hessels (2011) found that researchers' experience in firms and governments increases the likelihood that they will engage in interdisciplinary collaborations while it decreases the likelihood of mono-disciplinary collaborations. Similarly Carayol and Thi (2005, p. 77) reported that connections with industry is strongly correlated with interdisciplinary research.

There is also a perception that over time research is becoming more interdisciplinary (as shown by Porter and Rafols, 2009) in order to respond to increasing pressures to respond to social needs. In this sense, Gibbons et al. (1994) and Nowotny, Scott and Gibbons (2001) observed that research is undergoing a shift from a Mode-1 production of knowledge, which is mainly disciplinary and initiated by interests within academia, to a Mode-2 which is interdisciplinary, that displaces 'a culture of autonomy of science' (p.89) and addresses socially relevant issues. Other scholars have diagnosed a substantive transformation in science[2] along related lines (for a review see Hessels and van Lente, 2008). As Barry et al. (2008) noted, 'what is novel [in these analyses] is *the contemporary sense* that greater interdisciplinarity is a necessary response to *intensifying demands* that research should be integrated with society and the economy' (p. 23, italics are ours).

*Relationship between social relevance and local-issue research*

For our purposes, the key insight of Gibbons et al.'s analysis is the view that integration of science 'with society and the economy' is associated with the production of knowledge 'in the context of application'. This contextualization occurs not only at the organizational level (i.e. more interactions between producers and users of knowledge), but also in epistemological terms -- the knowledge that is produced in 'Mode 2' is now more attuned to specific places. This 'contextualized science', which involves the participation of a range of non-scientific stakeholders, produces research that yields more 'socially robust knowledge'. In this sense, Stiglitz (among others, such as Bones et al. 2011 and Gahi 2004), highlights that 'local researchers combining the knowledge of local conditions – including knowledge of local political and social structures -- ……provide the best prospects for deriving policies that both engender broad-based support and are effective…' (Stiglitz, p. 24 in Stone, 2000).

The importance of contextualizing of research in order to address to local social needs has long been recognized among scholars studying science and technology in developing countries (see review in Thomas, 2010). Alatas (1993, p. 312) proposed the 'indigenization of science' as a way for research to 'focus on problems more relevant to the Third World which have hitherto been neglected' and to move 'to specifying remedies, plans, and policies, and working with voluntary organizations and other non-governmental organizations, as well as with government in their implementation.'

The necessity to create, relate and adapt science to local contexts is more important in developing countries because dominant research agendas in international science (until recently mainly generated in rich nations) are more likely to respond to perspectives or problems posed in highly developed societies than to issues or problems of peripheral countries. Kreimer and Thomas observed that the production in developing countries of knowledge that is potentially applicable, but in practice is never applied --i.e. that it 'does not bring about product or process innovation, nor contribute to solving social or environmental problems' (cited in Thomas, 2010, p. 45).

---

[2] Although we use the generic term 'science' rather than science and technology, for the purposes of this article, we use it to mean the ensemble of institutions carrying scientific and technological research.

Local contextualization is more difficult to achieve in developing countries because shifting 'the research agenda toward creating the knowledge needed for solving local problems often involves isolation from the world community of scholars' (Sutz, 2003, p.56). Already in the late 1960s, Varsavsky exposed this 'tension of local scientists torn between international integration and the application of useful knowledge for society (cited in Kreimer, 2007, p. 2) and therefore creating the problem that springs from dependence on international science (Thomas, 2010, p. 40).

We contend that even 'global' problems such as climate change may sometimes require local data or the local contextualization of data (e.g. measurements of sea currents, winds, forests, etcetera. For instance an article related to climate change is 'Estimates of carbon reservoirs in high-altitude wetlands in the Colombian Andes').[3]

Since local contextualization is difficult to operationalize, in this article we investigate instead 'local-issue' research. While, in principle, research on local issues can be basic research and unrelated to social relevance, one can argue that it is more likely to be associated with a social or political issues (for instance, apparently basic biological research on ants in the Amazon forest may be related to environmental debates).

In summary, the literature suggests that socially relevant research is often associated with contextualization. In turn, it is therefore more likely to be associated with local-issues, especially in developing countries, where international research is less likely to be useful without some adaptation.

*The relationship between local-issue research and interdisciplinary research*

We furthermore contend that the literature also suggests that research addressing local contexts may often be interdisciplinary. Continuing the discussion of research in the context of application, Nowotny and Ziman (2002) state that.

> Practical contexts also have aspects that combine perspectives from different disciplines and are seldom intelligible without the development of novel inter-, multi- or transdisciplinary modes of knowledge production. (...) Localized science (...) is not just a 'perturbation' of the claims of universally valid paradigms or a denial of the feasibility of generalizing, reducing and deducing anything and everything. Knowledge production in the context of application is itself a fertile seedbed for the emergence of novelty. Localized investigations create genuine new knowledge. They can be full of surprises, especially when they combine knowledge elements from different realms, and mix them with societal expectations.

The claim that local-issue research should be interdisciplinary follows from the perception that local-issue research is needed to address socio-economic problems, particularly in non-Western contexts.

> Necessity and complexity have also been cited as reasons for IDR in and about developing countries. Shinichi Ichimura cautioned that the conceptual frameworks of traditional disciplines are often too narrow and too compartmentalized for the study of problems in other areas. Norman Dinges made a similar observation about cross-cultural research, suggesting interdisciplinary perspective grows as the 'indigenization' of research sensitive to local norms takes place; and Lawrence Murphy, using the example of the Social Research Center of the American University of Cairo (Egypt), has traced the movement from narrow, academically oriented research projects to more appropriate long-term

---

[3] This is one of the reported drivers of international collaboration with local partners (Wagner and Leydesdorff, 2004, as cited by Ordóñez-Matamoros, 2008, p.13): 'global problems requiring global solutions. Global warming would probably require research performed in different places of the planet to monitor and understand the cause.'

> interdisciplinary, multifaceted studies that analyzed problems of immediate concern to the host nation.
> (Klein, 1990, p. 45)

To conclude, there is a consensus in the literature that socially relevant research is most often interdisciplinary and some studies have shown link social relevance and local orientation. A portion of these scholarly works also attempts to weave both arguments together, arguing that interdisciplinary approaches are helpful to address local social needs. On these grounds, we formulate the hypothesis that there is a positive relationship between IDR and the production of local-issue knowledge.

Furthermore, given that solving local issues such as agricultural production demands knowledge from very different disciplines, we will also posit that the type of IDR required for local-issue research consists of the combination of distant disciplines such as atomic nuclear physics, oncology and sociology. This type of combination is what Yegros-Yegros et al. (2013) have called *distal interdisciplinarity*, as further discussed below. This stand in contrast to *proximal interdisciplinarity*, which is mainly focused on one discipline but takes some insights from neighbouring disciplines, for example a neuroscienace study that draws from related disciplinary categories such as physiology, pharmacology and clinical neurology.

## 3. Science policy and interdisciplinary research: the case of Colombia

IDR has received direct policy support in recent years, as reflected in reports by organisations such as the OECD (Godin, 2009), UNESCO (UNESCO, 2009), the U.S. National Science Foundation (Adams and Clemons, 2011: 218), the U.S. National Institutes of Health (Hall et al., 2008) among others (Brint, 2005). Yet, paradoxically, IDR remains discouraged in a variety of ways in many countries. For example, in universities, a prevailing 'silo' mentality also tends to suppress IDR, often indirectly as a result of promotion criteria and possibly enhanced by research assessment exercises that favour disciplinary approaches (see special issue edited by Laudel and Origi, 2006; Martin, 2011; also a review in Rafols et al., 2012).

This situation of explicit promotion of IDR accompanied with inadvertent suppression is seen as well in Colombia. Colombia is a country of approximately 45 million inhabitants which recently has been steeply increasing its number of scientific publications (Lemarchand, 2012, p. 294). As an upper middle income country, Colombia is making efforts to improve its S&T system and, as we show below, its science policy in the last two decades reveals that IDR has been promoted as a means to increase social relevance of research on local-issues, rendering it a good case for our study.

In Colombia, Colciencias is the organization that plays the lead role for the promotion and support of ST&I. Although originally created in 1968 mainly as a funding agency for research, it evolved into the central public organization for the formulation of national ST&I policy. IDR directed at socially relevant issues is explicitly promoted in the structure and operation of Colciencias, and this measure is reflected in its policies. For example, the organization has encouraged interdisciplinary collaboration between researchers, students and technicians among research groups.

Colciencias regularly issues open calls for problem-oriented projects, which in some cases are offered jointly with companies that require research in their field (oil and energy, for example). Other programmes that explicitly mention IDR are Centres of Excellence (interdisciplinary networks of groups based on strategic areas, Colciencias, 2004), Centres of Technological Development (private Industrial Technology Research Institutes), centres for agricultural research and other centres in cross-cutting technologies. The ambition of promoting IDR is also reflected in Government's strategic policy documents. For example, in 2000 the 'Departamento Nacional de Planeación' (National Planning Department) required explicitly that the Centres for Technological Development create interdisciplinary and inter-institutional innovation networks in order to propose and implement projects for technological improvement in Colombian firms (República de Colombia, 2000, p. 18). Also, in 2002, the National Development Plan of the Government included 'the strengthening of National Research Programs and their joint action articulated in complex topics and national priorities that require interdisciplinarity' (República de Colombia, 2002, 120).

At Colombian universities, which have been trying to develop their research capabilities, one can also find policies supporting IDR. For instance, the Universidad Nacional de Colombia (the largest public university) and the Universidad de los Andes (private) specifically mention support for IDR both in their mission statements and through calls for interdisciplinary projects (Universidad Nacional, 2005).

Despite these measures by government and universities, it remains unclear whether the implementation of these policies on collaboration is really fostering IDR practices, as illustrated by the assessment of research groups that is carried out regularly by Colciencias. This assessment ranks research groups in terms of bibliographic outputs that are based on disciplinary-centred method derived from publication patterns observed in physics that is being applied indiscriminately to all research groups, regardless of their area of research (Ruiz et. al., 2010; Restrepo and Villegas, 2007). As a result of an over-emphasis on the production of articles, researchers and universities participating in collaborative interdisciplinary groups, continue to focus on conducting disciplinary research (Chavarro et. al., 2010).

Colciencias has also acknowledged that it continues to operate through disciplinary lenses, for instance, in its internal structure for funding (discipline-based national programmes) and policy making. In 2004 there was a proposal to modify its internal structure to reflect a more socially relevant outlook (República de Colombia – Colciencias, 2004). Although it was not finally approved for reasons that remain unknown, the proposal illustrates Colciencias' awareness that a genuine modification of the organizational structure may be needed to achieve its stated goals for IDR, as noted above. While some initiatives have been developed, such as in encouraging the formation of collaborative interdisciplinary research groups, in practice, institutional inertia and operational practices remain important barriers to IDR.

These observations lead us to conjecture that the Colombian IDR policies in the main were, to date, declaratory, that is, the policies are mainly public statements without specifying the actions to be taken to implement their IDR policies. In summary, it is uncertain that the extent of IDR has been affected by these policies. It is worth noting, however, that recently regional authorities have been given a substantive budget for ST&I, drawn from taxes ('regalías') on the exploitation of non-renewable resources, such as minerals. This initiative is expected to have major effects on the local and socio-economic orientation of research, but it was introduced after we collected the data for this study and thus will not be reflected in this article. Moreover the effects may also be too early to capture.

## 4. Operationalization of interdisciplinarity and local issue orientation

We operationalize IDR and the production of local-issue research by drawing on publication data from journal articles, reviews and proceedings papers indexed by the Web of Science (WoS). First we chose the presence of the country name ('Colomb') in the abstract, titles or keywords as the criterion to identify locally oriented research, This approach was borrowed from a recent publication by Ordóñez-Matamoros, Cozzens and Garcia (2010).[4] Place-names act both as a coordinate system that locates geographically the action being performed and as a characterizing device that sets the action within a specific socio-economic context (for a conceptualization of place-names as indexical and characterizing signs, see Keates, 1996, pp. 81-82). Place-names 'are of such vital significance because they act so as to transform the sheer physical and geographical into something that is historically and socially experienced' (Tilley, 1994, p. 18).

---

[4] In a quick examination of the use of a 'place-name' we found that the percentage of publications that mention the country in their title, keywords or abstracts is much higher in Latin American countries, for instance, Colombia, Brazil, Argentina, Chile and Mexico, than in developed countries such as the U.S., the Netherlands, Germany and the UK. For the former group of countries, papers accounted for 15% to 25% of their total production, whereas for the developed countries the percentage was below 5%.

Second, following the National Academies (2005) we define IDR as the integration of knowledge and operationalize it through the use (i.e. integration) of bibliographic references from diverse disciplinary categories in one article. Then, we gauge the degree of interdisciplinarity using bibliometric indicators that measure the diversity of disciplinary categories in the references (Porter and Rafols, 2009), where diversity is computed taking into account the number, balance and disparity among the disciplinary categories (Stirling, 2007).

Third, we use a multivariate test to find whether there is a significant statistical relationship between degree of IDR in a publication and its local-orientation. We use two types of control variables:

(1) Degree of collaboration, given that collaborations tend to be more interdisciplinary (Qin et al., 1997) and that locally oriented research is likely to be more collaborative as well. We also check whether the collaboration is national or international, since one might expect that collaborations involving complex coordination (as it is the case in highly interdisciplinary projects) may become less likely as geographical distances increase. In other words, IDR collaboration requires significant coordination efforts, which may become more costly as the partners are further apart[5].

(2) Discipline of the publication, given that the degree of interdisciplinarity is highly dependent on disciplines (Porter and Rafols, 2009) and some disciplines such as ecology or public health are obviously more context-oriented than disciplines such as physics or computer sciences.

We run a logistic model with a composite measure of diversity first. Then, we unpack the various dimensions of diversity, which allows distinguishing distal versus proximal types of interdisciplinarity.

## 5. Data and methods

### 5.1. Data and sample

The dataset is comprised of articles, reviews and proceedings papers included in Thomson-Reuters' Web of Science (WoS) Database. These articles are authored by at least one researcher who was affiliated to a Colombian institution at the time of publication. We include records from 1991 (one year after the official foundation of the Colombian System of Science and Technology and the designation of Colciencias as the institution in charge of ST&I policy in the country) to 2010. All original data, analytical results and associated graphs are made available to readers in Supplementary File 1.[6] We only take into account records with more than three bibliographic references successfully categorized into WoS categories (this was necessary to construct a reliable measure of IDR). The application of these filters yielded 14,402 records, approximately 75% of the total sample of reviews, articles and proceedings papers published with Colombia in the period. Hence 25% of the publications were not used, in most cases due to the impossibility of classifying more than three references into WoS given the small number of references and that many of the journals referenced could not be classified.

### 5.2. Variables and methods

*Measure of local-issue orientation*

We define research orientation as 'local' when it directly mentions a word starting with 'Colomb' in the topic (title, abstract or keywords) and 'non-local' when it does not (1 means 'local' and 0 'non-local' orientation).

---

[5] We thank Lorenzo Cassi for suggesting this argument. The underlying hypothesis, derived from economic geography (Boschma, 2005; Ponds et al., 2007) is that collaborations involving different bodies of knowledge tend to be more geographically localized in order to compensate for difficulties in overcoming cognitive barriers.
[6] http://www.interdisciplinaryscience.net/pub_docs/idr-local-files/data_and_graphics(1).xls

In order to test the robustness of the method used as a proxy to identify local-issue research, two of the authors manually coded as locally oriented or not, two samples of 100 papers identified by the algorithm as local and non-local respectively. Articles which related to Colombian topics such as locally relevant diseases (such as Chagas), plants (such as oil palms) or related materials (such as fique fibers) were classified as local. The individual examination of each article involved making a dichotomous judgement of the degree of local orientation vs. the degree of 'universality'. Among the 100 articles classified as local, only 2 were perceived as non-locally oriented by both coders, and 10 by at least one examiner (false positives). Among the 100 articles classified as non-local, 9 were coded as locally oriented by both examiners and 26 by at least one examiner (false negatives). These results show that the classification of articles as local-issue research is problematic, but that the method used is an acceptable proxy for a large scale study such as this is (in the range of about 5-10% false positives and about 10-25% false negatives). Of course, such degree of error would not be acceptable for research assessments.

*Measures of interdisciplinarity*

The degree of interdisciplinarity of a publication is estimated by the diversity of WoS categories in its references, an indicator ranging from 0 to 1 (1 indicates totally interdisciplinary and 0 completely disciplinary). To do so, we follow Yegros-Yegros et al. (2013) (see also Rafols et al. 2012), who use each of the dimensions of diversity (variety, balance and disparity) separately as well as a synthetic measure of diversity (Rao-Stirling's) which combines all three dimensions (Stirling, 2007).

The equations for each measure of diversity are found below:

*Variety* = $v$ = Number of WoS categories

$$Balance = \frac{1}{\ln(v)} \sum_i p_i \ln p_i$$

$$Disparity = \frac{1}{v(v-1)} \sum_{i,j} d_{ij}$$ , sum only for those categories in the reference set.

$$Rao - Stirling\ Diversity = \sum_{i,j} p_i p_j d_{ij}$$

where $v_{max}$ = variety of the article with a greater number of WoS categories identified within the dataset, $p_i$ = proportion of elements in category i, $d_{ij}$ = distance between categories i and j (Rafols and Meyer, 2010, p. 267).

Each of the three first variables captures a different aspect of the general concept of diversity (Stirling, 2007, p. 710) while the fourth one is a synthetic formulation that takes into account all three other aspects (variety, balance and disparity). We should emphasize that there are other possible forms to operationalize the same properties.

*Variety* corresponds to the number of categories in which elements can be classified. *Balance* describes the evenness of the distribution of elements into categories. The form we use here is Shannon evenness. A sample is completely balanced if all categories share the same number of elements. *Disparity* is used to reflect the degree of the distinctiveness that exists between the elements of the distribution. If classifications are a means to separate elements, disparity is a relational property that tells the extent of separation (the distance) between the categories used. For example, soprano voices are closer to mezo-soprano than to contralto voices in terms of tone range. For this, a value for distance between elements (a metric) has to be set.

In our case, the measures of diversity are calculated for each article by classifying the bibliographic references into one or more WoS categories (generally representing academic subdisciplines), using the software Vantage Point.[7] An article will have high variety if it has references from many WoS categories. The article will have high balance if the proportion of references is evenly distributed across categories (e.g. 3 for Physical Chemistry, 3 for Applied Physics and 3 for Acoustics) and low balance if they are unevenly distributed (7 for Physical Chemistry, 1 for Applied Physics and 1 for Acoustics). The article will have a low disparity if the WoS categories in its references are cognitively related (e.g. Physical Chemistry, Applied Physics and Acoustics), and high disparity if the categories are distant in cognitive terms. The cognitive distances $d_{i,j}$ between categories are drawn from the metrics underlying the global maps of science done by Rafols et al. (2010) on journals in the WoS for 222 WoS categories (formerly Subject Categories) in 2007.[8]

*Rao-Stirling* diversity (also known as 'quadratic entropy') captures these three dimensions into a single indicator (Rafols and Meyer, 2010). It is not, however, a composite indicator resulting from a weighted sum of variety, balance and disparity --but a mathematical formulation that takes into account the three dimensions. The key advantage of this measure is that it not only takes into account the variety and balance of references across disciplinary categories, but crucially also considers how cognitively distant these categories are (i.e. the disparity). Intuitively, this means that a publication with references from atomic physics and cell biology is weighted as more interdisciplinary than one with references from cell biology and biochemistry.

The use of various measures of diversity rather than a single synthetic measure will allow us to explore the different effects that each of them (variety, balance, disparity) has on the propensity to conduct research on local issues. For the sake of parsimony, in the final conceptual analysis we will simplify the many potential types of IDR that can be obtained by combining high/low variety/balance/disparity into just two 'ideal' types (Yegros et al., 2013). On the one hand, we will call *distal interdisciplinarity* research building on relatively high balance and high disparity. This is research engaging distant disciplines such as physics, cell biology and sociology. On the other hand, *proximal interdisciplinarity* involves a relatively higher variety with lower balance and disparity. This would be research with a strong disciplinary core building on related knowledge.

The attribution of references to WoS categories is very problematic. It involves first the identification of the journal of a reference, and second its assignation to a WoS category(only possible if the journal is indexed in WoS). This assignation has been shown to be inaccurate – there is up to 50% disagreement between alternative classifications (Rafols and Leydesdorff, 2009, p. 1828). As a result, the diversity measure of a single article has a large noise and is not reliable. However the robustness of global science maps suggests that the error is not systematic and one can obtain good approximations with large numbers, (Rafols and Leydesdorff, 2009, p. 1829). As our sample consists of 14,402 publications, we are confident that the aggregation will yield reliable results. After classifying the references, a procedure in the statistical language R[9] was run on a list of articles to compute the indicators. These scripts in R are available in Supplementary File package 3.[10]

*Control variables*

In addition, we incorporated two control variables that may have effects on the relationship: (i) Collaboration and (ii) Discipline to which an article is more likely to belong, for instance Biosciences or Social Sciences. The variable Collaboration is a dummy variable with the categories International collaboration, National collaboration and No collaboration. This variable was identified from the field 'C1' in the WoS format, which holds the affiliation data of authors.

---

[7] www.thevantagepoint.com
[8] The similarity matrix between Web of Science Categories is available at Loet eydesdorff´s webpage for making overlay maps: http://www.leydesdorff.net/overlaytoolkit/ and at http://interdisciplinaryscience.net/pub_docs/idr-local-files/classification_of_journals.xlsx
[9] http://www.r-project.org/
[10] http://www.interdisciplinaryscience.net/pub_docs/idr-local-files/script/

The categorical variable for Discipline aims to control how the cognitive context may influence the local or non-local nature of the outcomes of research given that some disciplinary fields can be more prone to producing local studies than others (e.g. environmental studies tend to be more local than chemistry). The construction of this variable is based on the results of Rafols et al. (2010). Using factor-analysis, WoS categories were classified into 18 'Disciplines' according to similarity in citation patterns.[11] We assigned articles to the list of 18 disciplines by selecting the discipline with the highest number of references in a given article.

Table 1 shows a description of all the variables.

**5.3. Regression analysis**

To test the relationship between IDR and local research orientation, we used logistic regression. While other techniques, such as discriminant analysis, require meeting strict conditions of multivariate normality and equal distribution of variance and covariance matrices, logistic regression is robust when such conditions are not strictly met (Hair et. al., 2005, 276). For these reasons we have selected logistic regression using the statistical package SPSS.

The dependent variable is research orientation (that is, whether an article is Colombian or not), and the main predictor is the degree of interdisciplinarity, firstly as a synthetic variable (Rao-Stirling diversity) and secondly as represented by its different constituent dimensions (variety, balance, disparity). Hence, we performed the logistic regression in two blocks, first with Rao-Stirling as independent variable, second with the various diversity dimensions. We also tested for a possible inverted U-shape relationship between IDR variables and the dependent variable. The reduction in the -2 log likelihood (the variance) of each model is used as a criterion to assess the improvement in each block. We use three Pseudo-$R^2$ measures to assess the adequacy of the models. The first measure is Hosmer and Lemeshow's $R^2$, the second Cox and Snell's $R^2$ and the third Nagelkerke's $R^2$. These measures calculate the variation that is explained by the model based in -2 LL. The first is calculated as -2LL (new model)/-2LL (original model). 0 means 'no improvement' and 1 means 'total fit of the model'. This measure, however, does not take into account the size of the sample. For that, Cox and Snell's $R^2$ is used. As this measure cannot reach the theoretical maximum of 1, the correction by Nagelkerke is used. These three statistics help to assess the goodness of fit of the model (Field, 2009: 269).

---

[11] Groupings of Web of Science Categories are available at: http://interdisciplinaryscience.net/pub_docs/idr-local-files/classification_of_journals.xlsx

**Table 1: Description of the variables used in the study**

| Name | Type | Values | Role | Description |
|---|---|---|---|---|
| **Research orientation** | **Categorical** | 1 = local<br>0 = non-local | **Dependent** | If an article has a word with 'Colomb' in the title, abstract or keywords, it is considered local |
| **Variety** | **Numerical** | Between 1 and 222 | **Independent** | Number of Web of Science categories cited by each article. |
| **Balance** | **Numerical** | Between 0 and 1 | **Independent** | Balance in terms of proportion of references in each Web of Science category cited by an article. |
| **Disparity** | **Numerical** | Between 0 and 1 | **Independent** | Average distance between the Web of Science categories cited by an article. Distances are given by cross-citations between Web of Science categories across all science. |
| **Rao-Stirling Diversity** | **Numerical** | Between 0 and 1 | **Independent** | This variable combines three properties of disciplinary diversity: variety, balance and disparity. |
| **No Collaboration** | **Dummy** | 0 or 1 | **Independent** | 1 if there is no collaboration (only one Colombian address) |
| **International Collaboration** | **Dummy** | 0 or 1 | **Independent** | 1 if the affiliations has at least one non-Colombian address |
| **National Collaboration** | **Dummy** | 0 or 1 | **Independent** | 1 if more there is more than one Colombian affiliation in the article |
| **Discipline** | **Dummy** | **Agricultural sci.**<br>**Biomedical sciences**<br>**Business and Mgmt.**<br>**Chemistry**<br>**Clinical medicine**<br>**Cognitive sciences**<br>**Computer sciences**<br>**Ecology**<br>**Economics & geogr.**<br>**Engineering**<br>**Environmental S&T**<br>**Geosciences**<br>**Health services**<br>**Infectious diseases**<br>**Materials sciences**<br>**Physics**<br>**Psychology**<br>**Social studies** | **Independent** | Each article is assigned to the Discipline in which it had more references. Each Discipline is an aggregation of WoS categories in terms of cross-citations made by Rafols et al. (2010). |

## 6. Results

Tables 2 and 3 present general descriptive values for each variable in this study. Other graphs and tables can be found in the annex (Supplementary File 1[12]). Table 2 shows that 25% of the articles explicitly are local, i.e. reference Colombia. In terms of collaborations, we observe that about 60% of Colombian articles in the WoS database contain international affiliations. Figure 1 shows that the percentage of articles focused on Colombia has slightly decreased from ~30% to ~25%.

Figure 2 provides a descriptive view of the relationship between Rao-Stirling diversity and local research orientation. In considering the distributions of locally focused publications (black columns) and non-local publications (grey columns) separately, we see that the proportion of local-issue publications (black) is higher for more interdisciplinary articles (that is, Rao-Stirling diversity above 0.5), while the proportion of non-local articles (grey) is higher for lesser interdisciplinary articles (that is, Rao-Stirling diversity below 0.5).

---
[12] Also available at http://interdisciplinaryscience.net/pub_docs/idr-local-files/

It is worth noting that most of the publications present a Rao-Stirling diversity score between 0.4 and 0.6, that is, they are moderately interdisciplinary. The distribution of the variable shows a normal curve, within acceptable ranges of kurtosis and skewness (+/- 1) (Bulmer, 1979, p. 63). Extreme cases like publications with very low (0.1) or very high (0.8) Rao-Stirling diversity are unusual. When exploring variety, balance and disparity in regard to research orientation we find that the share of local papers is slightly greater for higher degrees of variety, balance and disparity (see Figures 7, 8 and 9 in the annex in Supplementary File 2).

An examination of the titles of the top ten most interdisciplinary articles according to Rao-Stirling diversity illustrates the relationship between IDR and local issues. As it can be seen in Table 4 below, seven out of the top ten most interdisciplinary articles are classified as local and most of them focus on topics directly related to Colombian issues: malaria, fruits, management of agricultural biotechnology in Colombia, and transport. Given that local papers are 25% of the total set, one would have expected only 2 or 3 out of 10.The local paper that appears to be less related to location is the one about history, but since it is on the history of engineering education, it could be considered as being relevant to the country's technological development. The majority of the ten articles appear to involve socially relevant research, except the one on entropy.

Similarly, we analyzed the papers with Rao-Stirling diversity equal to zero (lowest degree of interdisciplinarity), i.e. 76 papers referencing only one WoS category. Of these, only six papers were classified as local. Since local papers are 25% of the total set, one would have expected 19 instead of six (i.e. 76 over 4). We also examined the ten papers with the lowest Rao-Stirling diversity above zero (see Table 5). As can be seen in the Table 6, the majority of these papers do not have a local focus, and the one that has a local focus (cystic fibrosis) is actually on a mutation that is very common in Caucasian ethnic groups, but is not as common in Amerindian populations.

**Table 2: Descriptive statistics of the categorical variables Research orientation, Collaboration and Discipline.**

|  | Frequency | % |
|---|---|---|
| **Research orientation** |  |  |
| Non-Local | 10930 | 75.89% |
| Local | 3472 | 24.11% |
| **Collaboration** |  |  |
| National | 4968 | 34.50% |
| International | 8749 | 60.75% |
| No collaboration | 685 | 4.75% |
| **Discipline** |  |  |
| Agricultural Sciences | 997 | 6.92% |
| Biomedical Sciences | 2305 | 16.00% |
| Business and Management | 97 | 0.67% |
| Chemistry | 700 | 4.86% |
| Clinical Medicine | 1173 | 8.14% |
| Cognitive Sciences | 705 | 4.90% |
| Computer Science | 436 | 3.03% |
| Ecology | 1230 | 8.54% |
| Economics and Geography | 230 | 1.60% |
| Engineering | 439 | 3.05% |
| Environmental S&T | 615 | 4.27% |
| Geosciences | 334 | 2.32% |
| Health Services | 353 | 2.45% |
| Infectious Diseases | 1517 | 10.53% |
| Materials Science | 1808 | 12.55% |
| Physics | 1281 | 8.89% |
| Psychology | 158 | 1.10% |
| Social Studies | 24 | 0.17% |

**Table 3: Descriptive statistics of measures of interdisciplinarity**

|  | Minimum | Maximum | Mean | Std. Deviation |
|---|---|---|---|---|
| **Rao-Stirling Diversity** | 0.000 | 0.802 | 0.429 | 0.137 |
| **Variety** | 1 | 43 | 9.560 | 4.933 |
| **Balance** | 0.000 | 1.000 | 0.813 | 0.123 |
| **Disparity** | 0.000 | 0.999 | 0.629 | 0.125 |

**Figure 1. Percentage of publications with local focus over time**

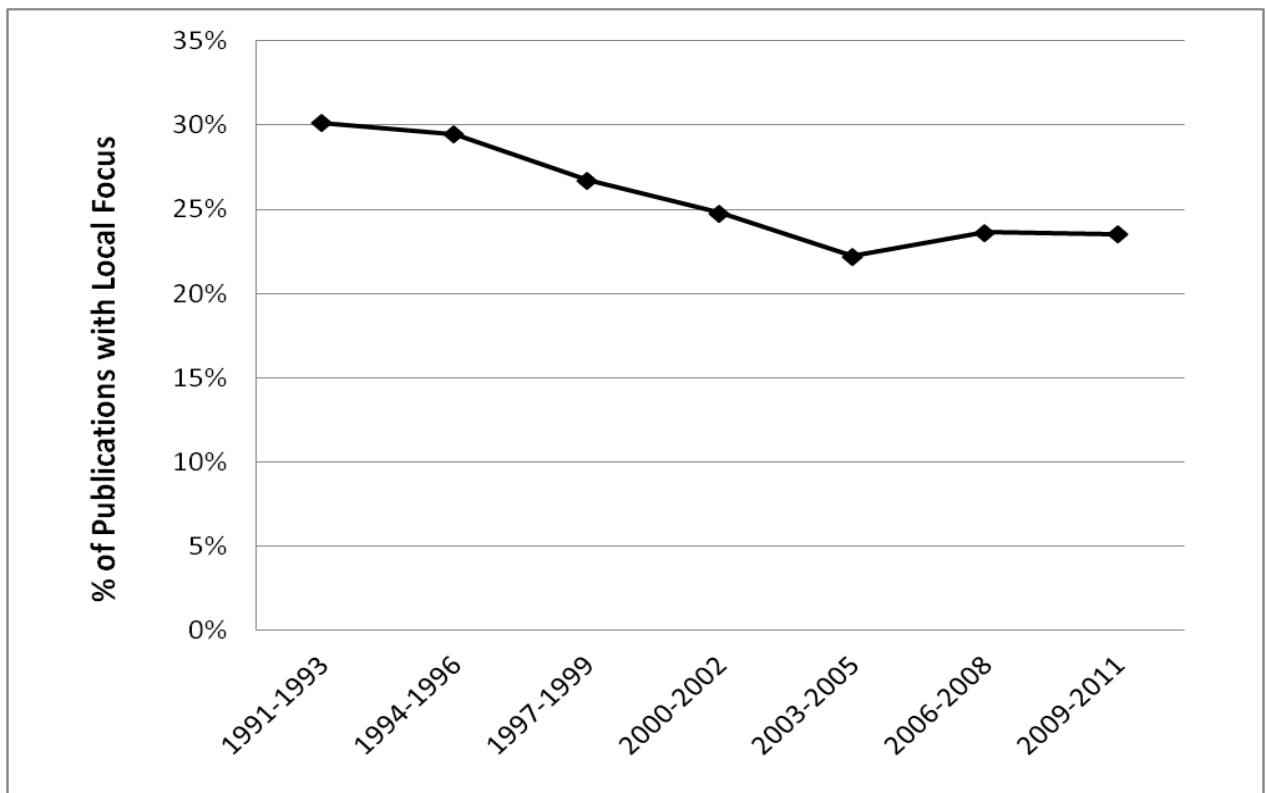

Note: The figure only includes publications used in this study.

**Figure 2. Percentage of local and non-local papers with a Colombian address by degree of interdisciplinarity (Rao-Stirling diversity)**

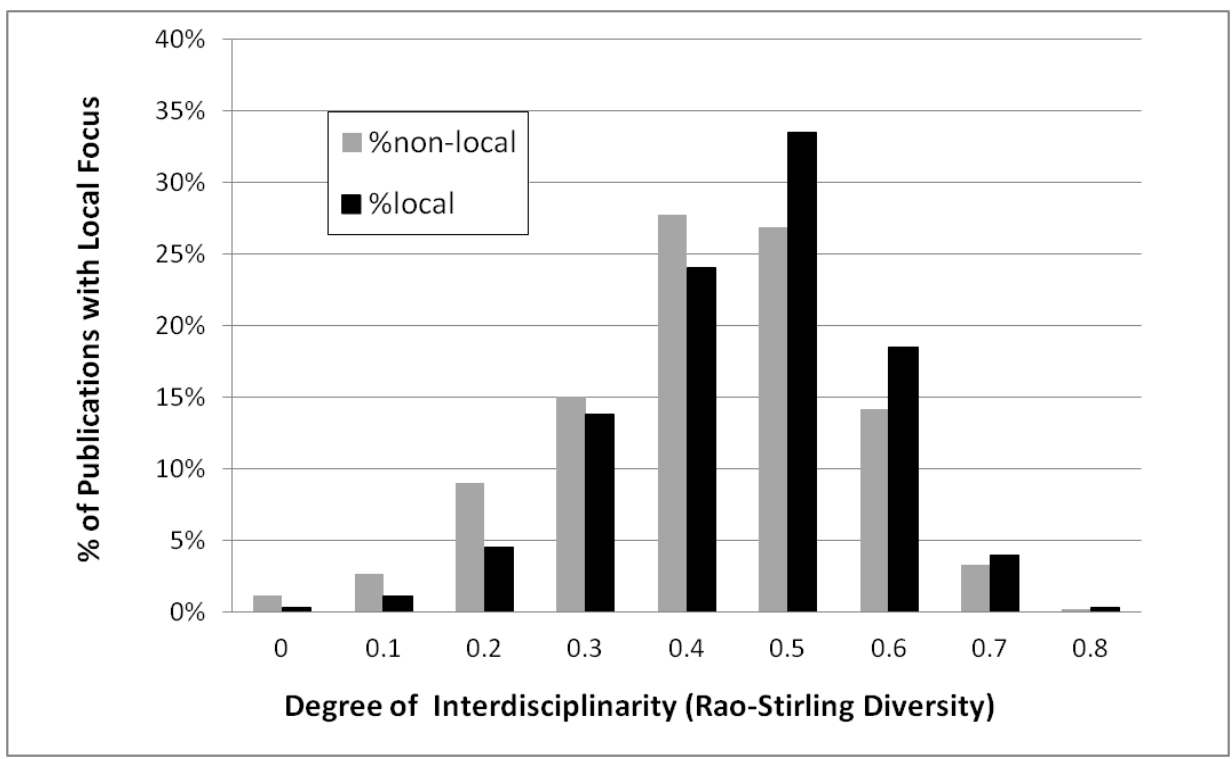

**Table 4. Top 10 most interdisciplinary articles in the set under study.**

| Title | Local | Rao-Stirling Diversity | Variety | Balance | Disparity |
|---|---|---|---|---|---|
| **Information and its management** for differentiation of agricultural products: The example of specialty **coffee** | Yes | 0.80 | 13 | 0.97 | 0.87 |
| A method for forecasting the seasonal dynamic of **malaria** in the municipalities of Colombia | Yes | 0.80 | 12 | 0.97 | 0.88 |
| A **transport network** reliability model for the efficient **assignment of resources** | Yes | 0.80 | 14 | 0.98 | 0.86 |
| Interpretation of **commercial production information**: A case study of lulo (Solanum quitoense), an under-researched **Andean fruit** | Yes | 0.80 | 26 | 0.93 | 0.88 |
| **Managing agricultural biotechnology** in Colombia | Yes | 0.78 | 26 | 0.93 | 0.86 |
| Analysis of **Andean blackberry (Rubus glaucus) production models** obtained by means of artificial neural networks exploiting information collected by small-scale growers in Colombia and publicly available meteorological data | Yes | 0.78 | 15 | 0.90 | 0.88 |
| **Engineering Education** and the Identities of Engineers in Colombia, 1887-1972 | Yes | 0.76 | 8 | 0.95 | 0.91 |
| Automatic **Detection of Pathological Voices** Using Complexity Measures, Noise Parameters, and Mel-Cepstral Coefficients | No | 0.79 | 26 | 0.91 | 0.83 |
| Using auxiliary information to adjust fuzzy membership functions for improved **mapping of soil qualities** | No | 0.79 | 18 | 0.93 | 0.87 |
| Entropy production in a radiating layer near equilibrium: Assaying its variational properties | No | 0.77 | 11 | 0.96 | 0.83 |

Note: Higher measures of Rao-Stirling (i.e. closer to one) indicate more diversity. A value of variety of 26 indicates that a publication has references in 26 out of the 222 WoS categories. Higher balance shows more evenness in the distribution of references. Higher disparity indicates larger cognitive distance between the references.

**Table 5. Articles in the lowest range of interdisciplinarity with Rao-Stirling higher than zero.**

| Title | Local | Rao-Stirling Diversity | Variety | Balance | Disparity |
|---|---|---|---|---|---|
| Effects Of Hydrostatic Pressure And Applied Electric Fields On The Exciton States In Gaas-(Ga,Al)As Quantum Wells | No | 0.023 | 2 | 0.52 | 0.11 |
| Application Of Sizing Design Optimization To Position And Velocity Synthesis In Four Bar Linkage | No | 0.023 | 2 | 0.41 | 0.15 |
| CFTR Mutations In Three **Latin American** Countries | Yes | 0.023 | 2 | 0.39 | 0.16 |
| Lp Estimations For The Class Of Pseudo-Differential Operators During Weyl-Hormander Calculations | No | 0.023 | 2 | 0.39 | 0.16 |
| Quasi-Two-Dimensional Magnetic Polaron: An Exact Self-Consistent Approach | No | 0.020 | 2 | 0.47 | 0.11 |
| An E-Based Mixed Formulation For A Time-Dependent Eddy Current Problem | No | 0.020 | 2 | 0.35 | 0.16 |
| Effects Of Growth-Direction Electric And Magnetic Fields On Excitons In Gaas-Ga1-Xalxas Coupled Double Quantum Wells | No | 0.019 | 2 | 0.30 | 0.19 |
| Superconducting, Surface And Interface Properties Of Ho(123) And Bi(2212) Films On Sapphire With Cerium Oxide Buffer Layers | No | 0.017 | 2 | 0.41 | 0.11 |
| Spin Evolution Of Accreting Young Stars. I. Effect Of Magnetic Star-Disk Coupling | No | 0.013 | 2 | 0.08 | 0.65 |
| Normalized Potentials Of Minimal Surfaces In Spheres | No | 0.013 | 2 | 0.25 | 0.16 |

## Logistic regression

We performed the logistic regression in two blocks. In the first block, we investigated the influence of Rao-Stirling diversity, with Collaboration and Discipline as controls. In the second block, we replaced Rao-Stirling diversity with the set of separate characteristics: Variety, Balance and Disparity. We tested the robustness of the results running again the models after removing potential outliers (standardized residuals > 3.0 or < -3.0), without a significant improvement of accuracy (below 2%). Table 6 presents the results of the regression:

**Table 6. Coefficients of the logistic regression**

| Variables | Model 1 | Model 2 |
|---|---|---|
| Rao-Stirling Diversity | 0.539 (1.715) ** | |
| Variety | | -0.257 (0.945) *** |
| Balance | | 1.051 (2.861) *** |
| Disparity | | 1.110 (3.034) *** |
| **Controls** | | |
| National Collaboration | 0.743 (2.101) *** | 0.770 (2.161) *** |
| International Collaboration | 0.155 (1.168) | 0.227 (1.255) * |
| **Disciplines** | | |
| Agricultural Sciences | 0.119 (1.126) | -0.025 (0.976) |
| Business and Management | 0.502 (1.653) * | 0.263 (1.301) |
| Chemistry | -1.925 (0.146) *** | -2.104 (0.122) *** |
| Clinical Medicine | -0.181 (0.834) * | -0.320 (0.726) *** |
| Cognitive Sciences | -0.187 (0.829) | -0.259 (0.771) * |
| Computer Science | -1.647 (0.193) *** | -1.943 (0.143) *** |
| Ecology | 1.195 (3.305) *** | 1.083 (2.955) *** |
| Economics and Geography | 0.212 (1.236) | -0.067 (0.935) |
| Engineering | -2.291 (0.101) *** | -2.610 (0.074) *** |
| Environmental ST | -0.290 (0.748) ** | -0.504 (0.604) *** |
| Geoscience | 1.805 (6.079) *** | 1.619 (5.047) *** |
| Health Services | 1.409 (4.093) *** | 1.249 (3.487) *** |
| Infectious Diseases | 0.586 (1.797) *** | 0.589 (1.802) *** |
| Materials Science | -2.891 (0.056) *** | -3.076 (0.046) *** |
| Physics | -4.406 (0.012) *** | -4.675 (0.009) *** |
| Psychology | 0.397 (1.487) * | 0.291 (1.338) |
| Social Studies | 0.956 (2.602) * | 0.746 (2.109) |
| **Constant** | -1.627 | -2.341 |
| | | |
| Cox and Snell's R2 | 0.199 | 0.207 |
| Negelkerke's R2 | 0.297 | 0.309 |

Note: Odds ratios are shown in parentheses. Model 1 includes Rao-Stirling diversity as a single measure for IDR. Model 2 replaces Rao-Stirling diversity with Variety, Balance and Disparity. The reference category for Collaboration is 'No Collaboration´. The reference category for Discipline is 'Biomedical Sciences'. Collinearity tests and correlations can be found in the annex (Supplementary File 2).
*** $p < .001$, ** $p < 0.01$, * $p < 0.05$

The logistic regression shows that IDR variables (Rao-Stirling diversity --Variety, Balance and Disparity) are significantly related to the production of knowledge on local issues. The odds ratio shows that for each unit increase in Rao-Stirling diversity (allowing for Collaboration and Discipline) it is 1.7 times more likely that an article is related to local issues. Disparity and balance also exhibit a positive relationship. A unit increase in these variables makes it approximately *three times more likely* that a paper is on local issues. Variety, on the other hand, contributes negatively to this relationship. A unit increase in variety makes it *0.9 times less likely* that a paper addresses local issues. Note that this decrease is only achieved when all other variables (including balance and disparity) are kept constant. In a descriptive mode, since more variety (more WoS categories cited) will be associated with increases in balance and disparity, which might make the paper more likely to be local, more variety leads to a greater probability that it will address local issues. Hence, the regression model is crucial for the observation. In short, the fact that the three aspects of diversity variables are mildly correlated (0.19 correlation between variety and balance, 0.37 between variety and disparity, and 0.19 between balance and disparity) explains why the overall effect of interdisciplinarity (Rao-Stirling diversity) has a positive effect in spite of the negative influence of variety.

The positive effect of disparity and balance on local-issue research suggests the specific type of interdisciplinarity that matters most for tackling local issues: research that bridges across large cognitive distances and that engages significant proportions of distant disciplines. As already explained above, following Yegros-Yegros et al. (2013), we call this *distal interdisciplinarity*. On the other hand, the negative effect of variety suggests that research that builds on many related sub-disciplines but has relatively lower disparity and balance (*proximal interdisciplinarity* as also referred to above) tends to be less related to local problems.

Third, it is important to note that the controls used in this analysis also have significant effects on the predicted variable. All collaborations are positively related to the production of knowledge on local issues. National Collaboration increases about two-fold the probabilities to publish on local issues, while International Collaboration by 1.2 times as compared to single-authored publications. In a similar investigation using teams (rather than articles) as units of analysis, and including a wider set of documents rather than only WoS publications, Ordóñez-Matamoros et al. (2010) found that 'a team's odds of involving Colombia in its research process are 2.2 times larger for those co-authoring with a partner located overseas than for those working individually or in collaboration with local partners' (p. 426).

We suggest that the difference in the findings may be due to the distinction, not made by Ordóñez-Matamoros, between papers of groups 'working individually' (no collaboration) and those 'in collaboration with local partners' (National Collaboration). From both studies one concludes that while international collaborations tend to increase the likelihood of tackling local issues, they do so less than national collaborations. This result is consistent with the argument, hypothesized using an economics geography framework (Boschma, 2005; Frenken et al., 2010) that research involving more coordination efforts is more likely to be carried out in relative geographical proximity. However, no conclusion can be made regarding the so-called 'outsourcing argument', namely that collaborating with international partners might shift the focus of attention away from local-issues (Ordóñez-Matamoros, 2008, p. 43).

The relationship between discipline and the production of knowledge on local issues also depends on the specific discipline. As compared to Biosciences (used as the reference category), there are some disciplines that increase the probability of producing publications on local issues. These are Business and Management, Ecology, Geosciences, Health Services, Infectious Diseases, Psychology, and Social Studies. Their odds ratios show an increase in odds between two (Social Studies) and five (Geosciences).

Finally, we tested for inverted U-shape relationships in each of the IDR-related variables. None of the quadratic variables showed a significant coefficient ($p < 0.05$). There is no evidence of an 'optimum' level of IDR after which the relationship changes its direction.

# 7. Discussion and conclusions

This paper has explored the relationship between IDR and the production of knowledge related to local issues. By using the case of Colombia (based on publication data extracted from the WoS), we have found that IDR publications tend to address local issues more often than disciplinary publications do. As discussed in the Introduction, this result is consistent with the perception in policy documents of the interrelation between local-issue research the social relevance of research and IDR.

Care, however, needs to be exercised about the interpretation of the regression model. What we show is a statistical association between local-issue orientation and the degree of interdisciplinarity. We do not intend to mean that there is a direct and causal relationship between interdisciplinarity and local-issue -- there is only an *increased likelihood* of local-issue research to be conducted through interdisciplinary approaches. Note that in Figure 2, there are many non-local issue research publications with IDR levels similar to those with local-issue research. Therefore, the statistical relationship we have found is not appropriate for evaluation purposes.

This relationship IDR and local focus might be explained by the fact that research related to local issues often aims to tackle or address specific, contextualized problems, and tends to be associated with socially relevant research (again, without implying a direct relationship). Socially relevant research as well often requires the mobilization and integration of diverse types of knowledge (Zierhofer and Burger, 2007; Rijnsoever and Hessels, 2011, D'Este et al., 2012), and this cognitive diversity is associated with interdisciplinary approaches (Rafols and Meyer, 2010). It follows that articles on local issues will tend to be more interdisciplinary as a result of their tendency to be more socially relevant. An inspection of the titles of the most interdisciplinary articles of the sample (see Table 4 above) supports this hypothesis. They are related, for example, to health (malaria), transport networks and agriculture (for example, the fruits *lulo* and Andean blackberry).

Our findings also reveal the specific type of interdisciplinarity that tends to be relevant to local issues. We find that articles with a focus on local issues tend to have a more balanced composition of highly disparate bodies of knowledge (more balance and disparity) in their references. An interpretation of these results is that local-issue research is associated with distal interdisciplinarity, which can be thought of as more difficult to achieve given the efforts required for combining disparate bodies of knowledge. For example, the study looking into the seasonal dynamics of malaria (Table 4 above) is based on insights from public health research, ecological dynamics, and statistical physics modelling. Our results also show, in contrast to distal interdisciplinarity, that higher number of disciplinary categories (high variety) is associated with less engagement with local issues (holding other variables constant). This suggests another type of IDR, proximal interdisciplinarity, which has a clear disciplinary focus with some, but limited, engagement (lower balance) with neighbouring disciplines (lower disparity). Proximal interdisciplinarity is possibly a more common approach in many studies largely because it is easier for researchers to communicate across short cognitive distances. Our study suggests that it is a form of IDR that is less likely to be related to local-issue research.

Our investigation, however, has some methodological limitations. First, different results might be found in high income countries in which the local focus is very likely not to be as evident from bibliometric measures as in a developing country such as Colombia. However, we think that our results could be generalized to other developing countries, in the so-called 'periphery' of the ST&I system. These countries are aspiring to participate in the global scientific community, while at the same time, they are trying to adapt and develop knowledge relevant to their local contexts with the aim of appropriating the socio-economic returns of S&T. Second, we use only publications indexed by the WoS, leaving out many articles by written by Colombian authors in Spanish and/or published Colombian journals and produced in national co-authorship (Salazar-Acosta et al., 2013). These articles are likely to be relevant regarding local-issue orientation. This gap is particularly problematic for the social sciences, which are very under-represented in the WoS. Third, the study uses a measure of interdisciplinarity that relies on the classification of references into WoS categories. Given that the classification of articles into WoS categories is very inaccurate (Rafols and Leydesdorff, 2009) and the number of references in an article is not very high, the measure used is very noisy, that is, it is likely to have variations due to contingent choices in reference selection. Nevertheless, we think that our sample is sufficiently large for the distribution to have a small standard error.[13]

Finally, we note that the findings of this article stand in contrast to those by Yegros-Yegros et al. (2013) who analyzed the relationship between IDR and citation performance. Yegros-Yegros et al. find a positive influence of variety and a negative influence of disparity and balance on the number of citations per paper. Since the authors' findings show the exact opposite effects of our findings, we speculate that related relations may be at play: (1) socially relevant research tends to be associated with cognitively disparate IDR (distal interdisciplinarity); and (2) socially relevant research (which is related to local-issue research), tends to be less valued in academic arenas (less cited) –therefore distal interdisciplinary papers gets less citations. These contrasts in the finding merit further exploration.

Drawing together our findings with those of Yegros-Yegros et al. (2013), we suggest that research assessment exercises that aim for 'high impact' in terms of citation counts (and possibly in journal ranking as well) may have the likely perverse consequence of sacrificing IDR that could produce local-issue research, which in turn could jeopardize the development of local S&T capabilities. Stated differently, by focussing on improving scientific 'excellence' using bibliometric measures such as citations in a developing country context, science managers run the risk of fostering the de-localization and decreasing the social relevance of research,

Our results are in agreement with the widely held view that supporting interdisciplinary research may be one of the measures which can help scientists to consider or engage with its immediate socio-economic environment, especially in developing countries (Macilwain, 2014). The policy analysis in Section 3 suggests that, although policies for promotion of IDR are in place (e.g. fostering heterogeneous collaboration within and beyond academia), institutional barriers still exert a strong pull towards disciplinary order (e.g. in evaluation and impact assessment exercises). Therefore, one may plausibly suggest stronger measures fostering IDR, such as promotion of collaborations, PhD studentships for IDR topics, etcetera, particularly for the production of knowledge related to local issues.

To conclude and to reiterate, this is an exploratory study and therefore one should be cautious about drawing policy implications. The study revealed a higher likelihood that research is locally oriented if research is interdisciplinary, but the difference between the lowest and highest IDR just doubles or trebles the likelihood that research is local. It is possible that with finer measures the results would be stronger. But there is also the possibility that the concepts used, IDR or local-issue research, are not sufficiently fine-grained to capture the underlying motivation for this investigation, namely, understanding how changing the way research is conducted is associated with its local social relevance. In a variety of studies it has been found that 'from an epistemological point of view, TDR [transdisciplinary research] does not represent a specific mode of knowledge production, but a rather heterogeneous conglomeration of different research activities' (Zierhofer and Burger, 2007, p. 51). This suggests that interdisciplinary research may also be associated with distinct ways of investigating, with different dynamics of social and local orientation. We hope that this study will stimulate further investigation of more refined variables for capturing local contextualization and associated research practices that support social relevance.

---

[13] An article-level classification system might provide a more accurate means of measuring the degree of interdisciplinarity (Waldman and van Eck, 2012), but the cognitive distances derived from article-based classification will require validation, whereas the ones we use here are known to be imprecise but have been validated in various studies at sufficient levels of aggregations (e.g. Soós and Kampis, 2011, Rafols et al., 2012).

**Acknowledgements**

We are grateful to Pablo d'Este, Alfredo Yegros-Yegros, Lorenzo Cassi and Alan Porter for fruitful discussions. We gratefully acknowledge support from the US National Science Foundation (Award #1064146 - 'Revealing Innovation Pathways: Hybrid Science Maps for Technology Assessment and Foresight'). The findings and observations contained in this paper are those of the authors and do not necessarily reflect the views of the National Science Foundation.

**Supplementary Files**

In order for readers to gain a deeper understanding of the material used in this paper, we provide further data and methodological details in the Supplementary Files.
- **Supplementary File 1** provides the data on Colombia used in the analysis
  http://interdisciplinaryscience.net/pub_docs/idr-local-files/data_and_graphics%281%29.xls
- **Supplementary File 2** is an annex that provides further details of the quantitative analyses, including descriptive statistics and support of the robustness of the regression.
  http://interdisciplinaryscience.net/pub_docs/idr-local-files/annex.docx
- **Supplementary File Package 3** provides the script and the baseline data for the computation of the diversity measures, including the distance metrics between Web of Science Categories
  http://interdisciplinaryscience.net/pub_docs/idr-local-files/script/